\documentclass[a4paper]{article}
 
\usepackage{amssymb}
\usepackage{amsmath}
\usepackage{latexsym}

\usepackage{shadow}

\usepackage{hyperref}

\usepackage[toc, page]{appendix}
\usepackage{appendix}
\usepackage{url}
\usepackage[all]{xy}

\def\etc {{\it etc.}}

 \def \d  {\rm{d}}
 \def \D  {\rm{D}}
 \def \ELequ  {Euler-Lagrange equation~}

\def\spt {space-time}
\def\dotc {{\dot{c}}}
\def\dotq {{\dot{q}}}

\def\R{{\rm I\!R}}
\def\vf  {vector-field}

\def\eg {{e.g.}}
\def\gr {general relativity}

\def\Hodge {{\star}}

\def\fb {fiber bundle}
\def\guill {\textquotedblleft ~}
 
\def\calA {{\cal A}}
\def\calD {{\cal D}}
\def\calF {{\cal F}}
\def\calJ {{\cal J}}

\def\calL {{\cal L}}
\def\calM {{\cal M}}

\def\calS {{\cal S}}

\def\dofs {degrees of freedom}

\def\QQ {{\bf Q}}
\def\Sect {\mbox{Sect}}

\def\Minks {Minkowski spacetime}

\def\VV {{\bf V}}
\def\andy{\mbox{~and~}}
\def\EQUIV {\Leftrightarrow}

\def\T{{\rm T}}
\def\coord {coordinate}
\def\Vol {\mbox{Vol}}
\def\undemi {\tfrac{1}{2}~}
\def\qdot {\dot{q}}

\def\mlr {Marc Lachi\`eze-Rey}

\def\Mk{{\rm I\!M}_k} 
\def\dmanif {differentiable  manifold}
\def\wrt  {w.r.t.~}
\def\ie {{i.e.}}
\def\eqbydef {~\stackrel{def}{=}~}

\newcommand{\dEL}[2] {\frac{\delta^{ \tiny  EL}  #1}{\delta #2}}
\newcommand{\der}[2]{\frac{{\mathrm d} #1}{{\mathrm d} #2}}

\newcommand{\pder}[2]{\frac{\partial#1}{\partial#2}}
\newcommand{\see}[1]{(see~ \ref{#1})}
\newcommand{\brak}[1]{\langle #1  \rangle}

\newcommand{\arXiv} [1] { \url{http://arxiv.org/abs/#1} }  

\newcommand{\blit}[1] {{  \emph{  #1}}}
\newcommand\innerp[1] { #1  \mathbin\lrcorner} 
\newcommand{\eql}[2]
{ \begin{equation} \label{#1} 
 #2
\end{equation}}

\title{Lagrangian Dynamics of Histories} 
\author{M. Lachi{\`e}ze-Rey, \\
 APC - Astroparticule et Cosmologie (UMR 7164) \\
 Universit\'e Paris 7 Denis Diderot
 10, rue Alice Domon et L\'eonie Duquet \\ F-75205 Paris Cedex 13, France\\
mlr@apc.univ-paris7.fr
}
\begin{document}
\maketitle
 
{\bf Abstract}

This paper presents (in its Lagrangian version)   a very general  \guill historicalÊ"  formalism for  dynamical systems, including  time-dynamics and  field theories. 
It is based on the universal notion of history.
Its  condensed  and universal  formulation  provides a  synthesis  and a  generalization  different  approaches of dynamics. It is  in our sense  closer to its  real essence.

The formalism is by construction explicitely covariant and does not require the introduction of time, or of a time function in relativistic theories.
It  considers    \spt ~(in field theories)   exactly in  the same manner than time in usual dynamics, with the only difference that it has 4     dimensions. Both time and \spt ~are considered as particular cases of the  general  notion of an \emph{evolution domain}.

In addition, the   formalism  encompasses  the  cases where  histories are  not functions 
(\eg, of time or of \spt),  
but forms. This applies  to electromagnetism  and to first order \gr ~(that we treat explicitely).
It has both Lagrangian and Hamiltonian versions. An interesting result is the existence of a covariant \emph{generalized symplectic form}, which generalizes the usual symplectic or  the multisymplectic form, and the symplectic currents. Its  conservation on shell  provides a genuine symplectic form on the space of solutions.

\section{Introduction}\label{intro} 
 
This paper  presents a general formulation of   Lagrangian  Dynamics, based on the notion of \emph{history}.
An history (for \guill kinematical history ") is a \emph{possible} evolution of a dynamical system. 
An history which obeys the dynamical equations is a \emph{physical} evolution (or particular solution, or dynamical history). 

Dynamics is usually   formulated in terms of  
\emph{dynamical variables} belonging to  a \emph{configuration space}.
Here we formulate it in terms of histories, \ie, possible evolutions of these variables.
Our calculations are thus performed in the \emph{space of histories}. It is infinite dimensional and one achievement is precisely to define such  calculus in such an infinite dimensional space.

This formalism describes
field theories  (hereafter, FT's) exactly in  the same manner  than the   (usual) time dynamics  (hereafter, tD), which appears  as a particular case. 
In time dynamics, 
an history is a function of time (assumed to be a well defined notion).  Most often, field theories (including the canonical approach to \gr) are also expressed \wrt time (or a \emph{time function}), although the price to pay is to loose covariance. Our approach respects covariance in considering that a field evolve not \wrt time, but \wrt \spt. Thus,    field theories  remain   entirely covariant and  find a very concise expression. It  is   analog to time evolution, with the  one-dimensional time  replaced by 4-dimensional \spt ~(this can be of course generalized to any number of dimensions). 
The difference  does not appear in the formalism, which  is designed in that purpose. Evolution is described  without    any notion of time, or any splitting of \spt. 

Moreover, it includes the case where the   histories (the fields)   are not \emph{functions}  like for a scalar field,  but \emph{r-forms}  on \spt. This applies to the    Faraday one--form in    electromagnetism,    or to the cotetrad and connection forms   in  first order  \gr.

We define a differential calculus in the infinite-dimensional space of \emph{historical maps},  defined as maps from the space of histories to itself. This allows us to perform a variational calculus in that space, and to  obtain an universal  dynamical (EL)  equation, with a very simple expression. It  applies,  in an entirely covariant manner,  to any  field  theories and usual equations are recovered as   particular cases.
We show (in the non--degenerate case) the existence   of a canonical \emph{generalized  symplectic form}  (it  reduces to the usual symplectic form in time dynamics),   which is  covariant and    conserved   on shell. 
To mention some general ideas underlying this approach, 
\begin{itemize}
  \item  dynamics is not defined versus time, but versus an \emph{evolution domain} which generalizes   the time line of tD. This is   \spt ~for FT's, where a space + time splitting is \emph{not} required, and where evolution is described  without any time-like parameter.
\item An history is not necessarily a function,  
but is generally  defined as a  [differential] form    on the evolution domain (or, more generally, as a section of some \fb ~based on it). This includes the   case of scalar fields, where functions are seen as zero-forms.
  \item A particular solution is an history which is  an orbit  of a dynamical  
  flow in the corresponding  bundle. This   flow is not one-dimensional like in time dynamics, but  has the   dimension    of the evolution domain.  It  may be called the \emph{general solution}. 
  \item Our formulation   holds 
in the space of histories which has   infinite dimension. Although this is not a manifold, we develop differential calculus in it, in the spirit of \emph{diffeology}\cite{PIZ}. 
  This work may   be seen   as a formalization and generalization    of \cite{Witten}. The corresponding Hamiltonian approach (not presented here,   see \cite{HHH}) leads to a synthesis between  that work and  the multisymplectic formalism.
\item We define, in the history space, a  canonical and covariant generalized  symplectic form. This is  a generalization of the usual symplectic form in tD, of the multisymplectic form and of the symplectic currents \cite{Witten} in  FTs. We show that it is conserved on shell. 
\end{itemize}
 
Our formalism (with its Hamiltonian counterpart, \cite{HHH})   offers   a   generalized synthesis between   the    multisymplectic geometry  (see, \eg, \cite{Helein}), the \guill covariant phase space "  approaches (see, \eg, \cite{Helein}),  the canonical approach and the geometry of the space of solutions. It remains entirely covariant.

The section \ref{intro}  introduces  the notion of histories  \ref{Histories}, in its general sense. It  defines their lifts (in the first jet bundle) to \emph{velocity-histories} involved in the Lagrangian dynamics. 
Section \ref{DynamicsL} introduces Dynamics in its Lagrangian version:  
the \ELequ leads to an universal historical  evolution equation. We  derive  the historical expression of Noether theorem. 
Section \ref{EX}  applies  to electromagnetism and to first order \gr.  

\subsection{Histories} \label{Histories} 

The central concept is that of \emph{history} (or possible motion, or kinematical,  or bare history). 
According to \cite{Sorkin},    histories \guill  furnish the raw material from which reality is constructed ".

As a general definition,      {\bf an history is  an  r-form on  an \blit{evolution  domain}  $\calD$},
 and    taking  its   values in a  \blit{  configuration space} $Q $ which represents the \dofs ~of the system: it is 
  a $Q$-valued form; a section of a    \blit{configuration bundle} $\QQ\to \calD$ which fiber $Q$ ~\footnote{An r-history may  also be seen as a map from $P^r(\calD)$, the space of r-paths  of $\calD$; such a point of view is convenient for a diffeological analysis \cite{PIZ}.}.  
An history (a field) may have components, in which it can   be  expanded. Each such component  is a scalar  r-form  over $\calD$, that we will write $c$. 
Without loss of generality, we treat the case of one history components (\ie scalar-valued r-histories)
to which we refer now as \guill histories ", and that we write   $c$. This corresponds to the case where $Q=\R$. An example of the general case is treated in    Section \ref{EX}.

   The   \emph{space of histories}   
 $\calS=\Sect(\QQ)$, or  a subspace of it (for mathematical conditions imposed on these maps, see, \eg, 
\cite{Barbero}). The  space of    physical motions (or particular solutions)   is a subspace of $\calS$: those   which obey  the motion equations, and it is an important task   of  dynamics  to select them \footnote{ particular solution may be   an equivalence class of such histories}.

 In   the usual  time dynamics (tD), the evolution domain is  the time line $ \simeq \R$ (or an interval of it). 
The non relativistic particle, for instance,  has   configuration space $\R ^3$  (particle position). Each   history (zero-history)  component
is  a \emph{function} over time: a zero-form    $c^i:~t\to c^i(t)$ (usually written $q^i$).  In  usual FTs, $\calD$  is the  \Minks ~$\Mk$ (or a more general \spt): for a scalar field, an history is a scalar function $c:~\Mk \to \R:~m \to c(m)$ ($c$ is usually written $ \varphi$).
  In electromagnetism, an  history is a   scalar value one-form $A$ (the potential) on \spt.
 Very generally, we define $\calD$ as  a $n$-dimensional manifold, possibly with a given metric.
(In tD, the existence of time is equivalent to that of a   metric $\d t \otimes \d t$ for the  evolution domain identified to the timeline). For field theories, $\calD$  is in general  a  metric  \spt ~but  conformal or  topological  theories involve no metric. In the case of  \gr,   the metric  is dynamical and  $\calD$ is a \dmanif  ~without  prior metric (see below). Our philosophy is  to treat $\calD$ as some kind of   \guill n-dimensional timeline~" \wrt which the evolution is expressed.

Thus an history (for history component) is    a scalar   r-form on   $\calD$, to which we refer as a r-history:
$$c\in  \Omega ^r _D\eqbydef \Sect[  \bigwedge ^r \T^*  \calD] \subset   \Omega  _D,$$
the space of  sections of the \fb ~$ \bigwedge  ^r \T ^* \calD  \subset \bigwedge   \T ^* \calD,$
(we note  $  \bigwedge   \T ^* \calD$ the bundle  of forms of all degrees;
we include \emph{functions} as the case $r=0$).
    
We emphasize the necessary distinction between  an  history, \ie, a section  of  the \fb ~$\QQ\to \calD$, that we always  write $c$, 
and its  possible values  (elements of  $Q$) that we write 
  $\varphi $   (usually  written $q$  in tD). 
    Working with $\varphi $ would  mean  working in    the finite-dimensional manifold of the  configuration bundle $\QQ$; working with $c$, as we do here,  means working in its space of sections. 
Despite the fact that this  is an  infinite dimensional space,  
we  define below some differential calculus on it and this a main idea of this paper. Our treatment is inspired by diffeological considerations  \cite{PIZ}.  Beyond its  compactness and generality, we claim that this is  closer to the physical reality.

 The philosophy of this paper is to transfer
the  (differential) calculus of configuration  space, or phase space,  to $\calS$, or to  other spaces with similar status.    
This  provides the possibility of a synthetic treatment   applying  both to tD and FT in a covariant way. This  offers in fact   a broader framework which  allows us to handle more general situations.

\subsection{Coordinates}
 
 It will be convenient, as an intermediary step for calculations   and possibly  as a pedagogical help, to   assume a system of [local]Ê \coord s on  $\calD$. They will  disappear in  our final results written in    covariant form. A choice of \coord s in $\calD$ generates adapted [local] \coord s in the various \fb s we will consider. 
 In tD, this is already provided by time $t$. 
 To treat all cases simultaneously, we refer to these \coord s as $x^\mu$. We write $\Vol$ the volume form  defined  from these  \coord s (when a metric is present, this is \emph{not} the volume form defined by it), and $\Hodge$ the corresponding    Hodge duality (those are non covariant entities).
We also write, as usual, $\Vol _{\mu} =\innerp{e_\mu}  \Vol = \Hodge (\d x ^\mu)$,

 $\Vol _{\mu \nu} =\innerp{e_\mu}  \Vol _\mu $, \etc
 
In  tD,  $\mu$ takes the only value  $ t$; $\Vol = \d t$,   $\Hodge \Vol=\Hodge (\d t) =1$,  and  $\Vol_t\eqbydef \innerp{\partial_t} \Vol =1$.
In FT's defined over 4-dimensional   \spt   ~$M$. In all the paper, we omit the wedge product sign for forms in  $\calD$, and 
 $\Vol =\epsilon_{\mu\nu\rho\sigma} ~\d x^ \mu~\d x^\nu~\d x^\rho~\d x^\sigma$ (not covariant). We will  use {multi-index} notations defined in \ref{multiindex}, so that
an r-history \eql{rhistory}{c=c_{\alpha_1...\alpha_r}~\d x ^{\alpha_1}...~\d x ^{\alpha_r}=
c_{\underline{ \alpha}}~\d \underline{ \alpha}.}

\subsection{Velocity-histories}

Any  section $c$ of $\QQ$ may be lifted \cite{SarletaWaeyaert}  (or \emph{prolungated}) to the 
first jet bundle  $J^1\QQ$ of $\QQ$. This gives, for  any history $c$, its  first jet extension $C\eqbydef (c, \d c)$ 
that we call      the corresponding   \blit{velocity-history}.
Here $\d$ is the  exterior derivative in~$\calD$;
  $\d c =
c_{,\mu}~\d x^\mu$, with  $c_{,\mu}\eqbydef \pder{c}{x^\mu}$, is a one-form on $\calD$.
\footnote{ This extends without difficulty to  $k^th$ order jets. 
}

Using \coord s, this becomes  $(c,  \dotc ) $ --- or $(q,\dotq)$ ---  in tD;  the multiplet 
$(c,  c_{,\mu}) $ --- or 
$(\varphi, \varphi_{,\mu}) $  --- in scalar field theories.
To be completely general, we may use
  multiindexes and represent   $C=(c,\d c)  $   as the multiplet  $(c_{\underline{ \alpha}}, c_{\underline{ \alpha},\mu})$  involving the [skew] components of the forms $c$ and $\d c$ 
\see{multiindex}.

 We consider  $C$ as a section of   the \emph{configuration-velocity  bundle} $\VV $, which identifies with  the  first jet bundle $J^1\QQ$ of $\QQ$
\footnote{
Strictly speaking,  $J^1 \QQ$ is defined as a \fb ~over $\QQ$. However it defines naturally a \fb ~over $\calD$. A velocity-history may be sen as a section of this bundle, which reduces to the tangent bundle in tD  (see, \eg, 
\cite{Gotay 1991,GotayIsenbergMarsdenMontgomery}).  }.
Its bundle manifold  is called the evolution space (Souriau).

We call
 $$  \calS_V =\Sect(\VV) \subset 
 \Omega^r_\calD  \times \Omega^{r+1}_\calD  \subset 
 \Omega_\calD  \times \Omega_\calD$$
    the space  of velocity-histories (technically, an \emph{exterior differential system}
\cite{CendraCapriotti}).
Since $j_1$ is canonical,   there is a   one-to-one correspondence between     $\calS_V$  and~$\calS$, and $C$ is nothing but a  more explicit      way to express $c$.

A first  idea of this paper is to express the (Lagrangian) dynamics in $\calS_V$ instead of $\VV$, \ie, in the space of sections (histories) rather than the bundle itself. This implies the replacement of functions by \emph{historical maps}, as they are defined below.

\section{Dynamics : action and Lagrangian}\label{DynamicsL}

The \blit{action}  is a map $\calA$ associating     to any history (velocity-history)   a real number.
It   is   expressed  as  the integral
   $$\calA:~C\to \calA[C] =\int _\calD ~Ê\calL(C) .$$  
     
Here  $Ê\calL(C)=\calL(c,\d c) $, to  be integrated on $\calD$, is  an $n$-form on $\calD$. The map  $\calL$   
associates,     to any velocity-history $C$, the    n-form $\calL(C)$ on  $\calD$.
We call it the    \blit{Lagrangian functional}, a specific case of  \emph{historical-maps}, defined below.

Let us make the link  with the usual    physicist's conception, \eg, for  a scalar field (zero-history) with a 
Lagrangian   \emph{scalar function}  
 \eql{Lagrangianscalarfunction}{\ell:~ \VV \to \R:~(\varphi,v_\mu)\to \ell(\varphi,v_\mu) }
  on the configuration  - velocity  manifold $\VV$. \footnote{The latter admits   \coord s $(x^\mu, \varphi,v_\mu)$ 
but covariance requirements imply   no explicit  dependence on the $x^\mu$.}
The n-form $\calL (c, \d c )$ is  defined through 
$$\calL (c, \d c )(x) =  \ell  [ c  (x), c_{, \mu}(x)]~\Vol.$$

This remains valid when an history is an r-form $c$,
excepted that  $C$ is then  expressed by components
  $(c_{ \alpha_1 ...\alpha_r},c_{  \alpha_1 ...\alpha_r, \mu})$  and 
$$\ell:~ \VV\to \R:~(\varphi_{ \alpha_1 ...\alpha_r},v_{ \alpha_1 ...\alpha_r~\mu} )\to \ell(\varphi_{ \alpha_1 ...\alpha_r},v_{ \alpha_1 ...\alpha_r~\mu})  .$$
 
\subsection{Historical maps}  

We define an \blit{historical map} (Hmap) as a generalization of the notion of  \emph{functional}:

We consider first   maps
$ \Omega_D\to\Omega_D $: each such map  takes  an r-form  over $\calD$
as argument   
and returns an R-form  over $\calD$ ($r$ and $R$ are integer $\le n$).
We generalize still further by allowing   \emph{two} arguments, so that we define  an     an \blit{historical map} (Hmap)  as a map 
 \footnote{ For a k-order theories,  $( \Omega_D) ^2$  would be generalized to $( \Omega_D) ^k$.}. \\
$$F:~\calM\eqbydef ( \Omega_D) ^2\to\Omega_D:~(c,\gamma)\to F(c,\gamma).$$

Occasionally we will call such a  Hmap  a Hform of type [0,R]:  the  0 refers to the fact that this is a map, that we consider as a  zero-form   (Hforms of higher degree will be considered later); and   $R$ refers to the grade of the values taken by  $F$.

The wedge product in $\calD$ defines a product  of the   Hmaps (that, again, we always write implicitly by simple juxtaposition):
$$(F~ G) (c,\gamma) \eqbydef F (c,\gamma) ~ G (c,\gamma).$$ 
 Thus    the Hmaps
form an    algebra  $\calF=\Omega^0(\calM)$, that we will treat    
like an algebra of functions   over $\calM$, that we treat itself as an infinite dimensional manifold.
We intend to  define differential calculus  in that space, allowing 
variational calculus. This may be seen as a generalization of the variational bicomplex of    \cite{BRIDGES}, 
or of   the double complex structure  introduced by  \cite{Deligne}, with the difference that we work in a space of sections rather than in a \fb.
Note also 
 that similar approaches   (\cite{Zuckerman}, \cite{Deligne})   consider  elements of $\Omega (\Sect(\Omega_D \times D)$.)

In  practice we will only consider maps from a certain   subset of $( \Omega_D) ^2$,  but we consider here the   general case. The  Lagrangian functional $\calL  $ above is a typical    [0,n]--Hmap. We will however only consider as    arguments a pair    $c$  (an r-history)  and   $\gamma=\d c$, the latter being  a (r+1)-form representing its exterior derivative. It returns an n-form.

First we notice    that the differential $\d$ on $\calD$ is easily lifted to $\calF$ through the formula 
$$(\d F) (c,\gamma) \eqbydef  \d (F(c,\gamma)) .$$ It improves the grade from [0,R] to [0,R+1]. We call it occasionally  \guill horizontal  derivative " but we do  not consider it as a genuine part of our  differential calculus on $\calF$  since it does not change the status of $F$ and does not    allow  variational calculus.
In that purpose,   we  introduce an second \guill vertical " external  exterior derivative $\D$, different from $\d $ and commuting with it.    
\cite{Khavkine}

  \subsection{Differential calculus for historical maps}

We first define the  two     basic   \emph{partial  derivative operators} $\partial _c=\pder{}{c}$ and $\partial _\gamma=\pder{}{ \gamma}$   acting on $\calF$ through the variation formula  
(wedge product in $\calD$ is always assumed) \eql{variationformula}{\delta F=\delta c~\pder{F}{ c} + \delta \gamma  ~\pder{F}{\gamma } .}

All quantities involved are Hmaps and their  arguments 
 $c$ and $\gamma$    play the role of   \guill  \coord s " in $\calM$ \footnote{
 $c$ and $\gamma$   may be themselves seen as particular (tautological)  {Hmaps} returning their own values.}.  The types of  $F$, 
$\pder{F}{ c}$ and  $\pder{F}{ \d c}$  are respectively [0,R], [0,R-r] and [0,R-r-1].
This gives a complete definition, but  we give more explicit expressions in Appendix.
 
We consider  
$\pder{ }{c}$ and $\pder{ }{\gamma}$  as basic  \vf s on $\calM$, and define   the  \blit{general  \vf}  as 
\eql{generalvf}{V=V^c~ \pder{ }{c}+ V^{\gamma}~\pder{ }{\gamma},}
whose   components $V^c$ and $   V^{ \gamma }  $  are  arbitrary Hmaps   themselves 
(wedge product  in $\calD$ still assumed).  
It acts  on an arbitrary  Hmap     $\beta$,   as 
$$V(\beta)=V^c ~Ê \pder{\beta}{c} + V^{\gamma } ~\pder{\beta}{\gamma}. $$
We write $\chi(\calM)$ for their set. 
The generalized multi-\vf ~is defined through antisymmetric tensor product. 

We   now define differential forms:   first    the    \blit{basis one-forms} $\D c$ and $\D  \gamma$  through their actions on   \vf s:
$$\brak{\D c,V  }=V^c;~~~~~~~~\brak{\D\gamma,V}    =V^{\gamma}.$$
And the general  one-form is  $\alpha=\alpha_c ~\D c+ \alpha_{\gamma}~    \D \gamma, $
with components  $ \alpha_{c}$ and $  \alpha_{\gamma} $ arbitrary Hmaps again.
We write $\Omega^1(\calM)$ for their set.  
We may  now define the (vertical) exterior derivative of an arbitrary Hmap $F$ as       \eql{DDD}{\D F=\D c~ \pder{F}{c} +\D\gamma ~ \pder{F}{\gamma } ,} 
which is now of type [1,R]. 
 
 The wedge product of forms,  $\wedge$ (not to be confused with the   wedge product on $\calD$ written by simple juxtaposition), is the  antisymmetrized tensor product, as usual. This defines $\Omega(\calM)$. 
 The table illustrates the action of the horizontal and vertical derivatives $\d$ and $\D$.

\begin{displaymath}
\xymatrix{
\Omega^0 (M ):& [0;R]    \ar[d]^{\D} & \ar[r ]^{\d}   & [0;R+1] \ar[d]^{\D}& &   \\
\Omega^1 (M ):& [1;R]       & \ar[r ]^{\d}   & [1;R+1]& &   \\
     }
  \end{displaymath}
 
Since we are interested in variational calculus, we express an arbitrary variation    as  resulting from the  action of a  \vf ~$\delta=\delta^c~\partial _c+  \delta^\gamma~\partial _\gamma$ on the arguments $c$ and $\gamma$, namely
\eql{HistVariation}{\delta c=\brak{\D c , \delta}=\delta^c,~~ \delta \gamma=\brak{\D \gamma , \delta}=\delta^\gamma.} This requires   $\delta ^c$ and $\delta ^\gamma$
to be of  the same  grades than $c$ and $\gamma$ respectively. 
It is easy to check that, with  \ref{DDD}, this leads to
\ref{variationformula}. In Lagrangian calculus, we only consider the case where $\gamma =\d c$, and thus we restrict to variations  obeying $\delta \gamma=\delta (\d c)=\d (\delta c)$.

\subsection{Historical momentum}

Applying this calculus to the Lagrangian gives 
\eql{momenta}{\D \calL=\D c~ \pder{\calL}{c} +\D Ê(\gamma) ~ \pder{\calL}{\gamma} .}
We call \blit{ historical momentum}
the  Hmap $P\eqbydef \pder{\calL}{\gamma}$.
It is of type
[0, n-r-1], 
with   arguments $c$ and $\gamma$.  In all calculations, $\gamma$ will always take the value~$\d c$.
The Hmap $P$  expands naturally in \emph{ dual components},  as  $P =P^{\underline{ \mu}}~\Vol_{\underline{ \mu}}$, where we use the polyindexes defined in \ref{multiindex}.
 This is  defined through 
$  P^{\underline{ \mu}}  (c,\gamma)= P(c,\gamma)^{\underline{ \mu}}$.  In time dynamics, $P$ identifies with  the usual momentum. For a scalar field,  the dual  components $P^\mu$   correspond to the  so called \blit{polymomenta}
 used  in the multisymplectic formalism. The cases of electromagnetism and \gr ~are treated below.
Similarly, one may call  $ \pder{\calL}{ c}$ the force.

\subsection{\ELequ  }

Lagrangian historical dynamics corresponds to the case where $\gamma=\d c$.  Then, since $\d $ and $\D$ commute, the previous identity takes the form 
\eql{momentaB}{\D \calL=\D c~(
\dEL{\calL}{c})- \d    \Theta.}
where we  defined the  Euler-Lagrange derivative \eql{ELform}{ \dEL{\cdot}{c}\eqbydef   \pder{\cdot}{ c } -  (-1)^{|c|}~\d \pder{\cdot}{(\d c)},}
with $|c|=${grade~of~ c}. 
We also define the [1; n-1 ]--Hform    (implicit wedge product in $\calD$)
\footnote{  in the sense of a 1-form for the vertical  calculus, and a (n-1)-form for the [horizontal]Ê  calculus in $\calD$; see Appendix for notation.}
$$\Theta\eqbydef - \D     c  ~P=\D c~\pder{\calL}{(\d c)} .$$

 It  gives by vertical   derivation the   [2; n-1]--Hform $\omega \eqbydef  \D \Theta =\D P \wedge \D c$ 
 (implicit wedge product in $\calD$). This is a closed two-form. When the Lagrangian is non singular, it  is non degenerate and  plays the role of  
 symplectic form, excepted that it take its values in $\Omega^{n-1}$ instead of $\R$.
 We    call $ \Theta $ and $\omega=\D \Theta $ the        \emph{historical Lagrangian  forms}; in the non--degenerate case,     the historical  symplectic potential and form. This is the \emph{historical version} of the usual Lagrangian  (or symplectic) forms  on  the velocity--configuration space  (see, \eg
\cite{Kharlamov,Cattaneo}). \footnote{ or of the pre-sympletic structure of the evolution space.}

An arbitrary variation of an history  is seen as the action   of a \vf ~$\delta$ in $\calF$  
as given by \ref{HistVariation}, such that $\delta^{\d c}=\d \delta^{ c}$.
This implies $$\delta \calL=\delta c~(
\dEL{\calL}{c})+ \d    (\delta c~\pder{\calL}{(\d \varphi)}).$$
Since the last term does not contribute to the action, stationarity corresponds to the \ELequ
\eql{ELEQU}{\dEL{\calL}{c}=0.}

This  equation applies    equally well to time dynamics  and field theories. In the latter case, 
it is  explicitely covariant. It    includes  the case where   $c$ is a r-history (a r-form)  rather than a  function, as  illustrated in the last section.

\subsubsection{On shell conservation}\label{On shell conservation}

 \emph{On shell},  
\ref{momentaB} implies $$\D \calL =- \d \Theta,$$ and thus
$$\D \D \calL=0 =-\D \d \Theta= -\d \D  \Theta= -\d \omega :$$ the generalized symplectic form is conserved on shell. 
 This is the covariant version   of the conservation of the symplectic current.

Since the value of $ {\omega}$ is a $(n-1)$-form on $\calD$, it can be integrated along  a $(n-1)$-dimensional  submanifold of $\calD$. This 
 provides a scalar-valued symplectic form on the space of solutions.
 On shell, the conservation of  $ {\omega}$   implies that this scalar form does not depend on the choice of the hypersurface (assumed time-like for FTs). This provides the  canonical (scalar valued) symplectic form  on the space of solutions  introduced by \cite{Witten}, so that our result may be seen as a generalization of their work.

\subsection{Symmetries}

A \vf ~$\delta$ is a symmetry generator  when it does not modifies the action. This  means that it modifies $\calL$ by an exact  form $\d X$ only. Hence, for a symmetry, $$\delta c~(\dEL{\calL}{c})-\d(\delta c~P)=\d X.$$ Defining the \emph{Noether  current} (three-form)  $j\eqbydef    X + \delta c~P$, we have the conservation law
$$\d j =  \dEL{\calL}{c}~\simeq~ 0 \mbox{~(on~shell)}.$$
Locally, $j=\d Q$, which defines the \emph{Noether charge density}   (n-2)--form $Q$ \cite{Wald}. 

A   diffeomorphism  of $\calD$ is obviously a symmetry  since in that case $\delta\calL =L_\zeta \calL = \d (\innerp{\zeta} \calL)$, where $\zeta$  is the generator.

\section{Examples}\label{EX}

\subsection{Time dynamics}

An history $c$ is a function of time: a zero-form on $\calD$ and   $\d c   =\dotc ~\d t $  is a 1-form. Writing $\ell$  the usual scalar Lagrangian function,
$$\calL(c,\d c)=L(c,\dotc)~\d t=\ell[c(t), \dotc(t)]~\d t $$ is a one-form. Thus   
$$\pder{\calL}{  c}= \pder{L}{c}~\d t= (  \pder{\ell}{q}\circ C)~\d t $$ and 
$$\pder{\calL}{\d c}= \pder{L}{\dotc} = \pder{\ell}{v}\circ C   $$ are respectively a 1-form and 
a 0-form.

The EL equations \ref{ELform} take  the usual form:
$$ (\pder{\ell}{q}\circ C)~\d t +\d (\pder{\ell}{v}\circ C  ) =
 [\pder{\ell}{q}\circ C  + \der{}{t}(\pder{\ell}{v}\circ C )]~\d t=0,$$
that physicists usually condense as $  \pder{\ell}{q} + \der{}{t}(\pder{\ell}{\qdot} ) =0$.

Note that  $P=P^t$ is a zero-form, $\Vol_\mu=\Vol_t=1$. Then, $\delta \calL=
\pder{\calL}{c}~\delta c+ P    ~\delta \dotc ~Ê\d t
=\pder{\calL}{c}~\delta c+P~\delta Ê(\d   c)  $:  a covariant expression  of tD.

\subsection{Electromagnetism in \Minks}

The dynamical variable is the one-form (1-history) $A$. The   Lagrangian functional $\calL=\undemi \d A ~(\Hodge \d A )$ (Hodge duality  in \Minks).
It results 
$\pder{\calL}{(\d A)}=  \Hodge \d A  $
and the \ELequ, 
 $$\d(\Hodge \d A  ) =0\EQUIV \Box A=0,$$ reduces to the Maxwell equation.

\subsection{ First order   gravity}

We   start  from the  Lagrangian functional (on a 4-dimensional differentiable  manifold $M$ without metric)
\eql{eomegaLAGR}{\calL   = \epsilon_{IJKL}~ e^I~e^J~ ( \d~\omega^{KL}+  (\omega\omega)^{KL}) .}
(Again, we   forget   the wedge product signs with the convention that  juxtaposed  forms on $M$    are wedge-multiplied). 
The dynamical variables are the cotetrad one-forms 
$ e^I$ and the Lorentz-connection one-forms  $\omega^{KL}$, with respective momenta 
$P_I$ and $\Pi_{KL}$.
We have 
\eql{Pi}{\calL_I\eqbydef \pder{\calL}{ e^I}=
2~ \epsilon_{IJKL}~ e^J~ ( \d~\omega^{KL}+  (\omega\omega)^{KL}) ~\andy~
P_I\eqbydef\pder{\calL}{\d e^I}=0;}  
\eql{deromega}{ \calL_{KL} \eqbydef\pder{\calL}{  \omega^{KL}}=2~ \epsilon_{IJNL}~ e^I~e^J~\omega^N_{~K} =2~ \epsilon_{IJKL}~ e^I~\omega^J_{~M}~e^M }
  (after some algebra) and
\eql{PI}{  \Pi _{KL}\eqbydef \pder{\calL}{\d \omega^{KL}}= \epsilon_{IJKL}~ e^I~e^J.}
  
  The EL equations are easily derived as 
  $$ \pder{\calL}{ e^I}= 2~ \epsilon_{IJKL}~ e^J~ ( \d~\omega^{KL}+  (\omega\omega)^{KL})=0 $$  
which    means zero Ricci curvature; and   
  $$\pder{\calL}{\omega^{KL}}+\d  \pder{\calL}{\d \omega^{KL}}=
     \epsilon_{IJKL}~ e^I~[\omega^J_{~M}~e^M+ ~\ de^J]=0$$
which means zero torsion.

\subsection{Lorentz invariance}

An infinitesimal Lorentz transformation transforms the tetrad and the connection as
$$\delta e^I= \lambda^I_{~J}~e^J;$$
$$\delta \omega^{IJ} =
\lambda^I_{~K}~\omega^{KJ}-  \omega^{IK}~\lambda_K^{~J}+\d  \lambda^{IJ}.$$

The Lorentz invariance of $\calL$ leads to the conserved Noether  current: the  3-form (on $\calD$)
$$J =\Pi _{IJ} ~\delta \omega^{IJ} 
= \Pi _{IJ} ~(\lambda^I_{~K}~\omega^{KJ}-  \omega^{IK}~\lambda_K^{~J}+\d  \lambda^{IJ}).$$

Conservation takes the form $$0= \d J = \d \Pi _{IJ} ~(\lambda^I_{~~K}~\omega^{KJ}-  \omega^{IK}~\lambda_K^{~J}+\d  \lambda^{IJ})
+  \Pi _{IJ} ~\d (\lambda^I_{~K}~\omega^{KJ}-  \omega^{IK}~\lambda_K^{~J})
.$$
This  splits as 
$$0= \d \Pi _{IJ} ~(\lambda^I_{~K}~\omega^{KJ}-  \omega^{IK}~\lambda_K^{~~J})
+  \Pi _{IJ} ~ (\lambda^I_{~K}~ \d \omega^{KJ}-  \d \omega^{IK}~\lambda_K^{~J});$$
$$0= \d \Pi _{IJ} ~( \d  \lambda^{IJ})
+  \Pi _{IJ} ~ (\d \lambda^I_{~K}~\omega^{KJ}-  \omega^{IK}~\d\lambda_K^{~~J}).$$

The first expresses the \emph{global} invariance as
$$0= \d \Pi _{IJ} ~ \lambda^I_{~K}~\omega^{KJ}- \d \Pi _{IJ} ~ \omega^{IK}~\lambda_K^{~J}
+  \Pi _{IJ} ~  \lambda^I_{~K}~ \d \omega^{KJ}-    \Pi _{IJ} ~\d \omega^{IK}~\lambda_K^{~~J}  $$
$$ =2~  \lambda^A_{~K}~ \d( \Pi _{AJ} ~\omega^{KJ} ).$$
Since this is for arbitrary $\lambda$, and taking account of antisymmetry, this implies (on shell)
$$\d \calJ_{AB}=0; ~
 \calJ_{AB}=   \Pi _{[AJ} ~\omega^{J}_{~~B]}=\epsilon   _{[AJMN} ~e^{MN}~\omega^{J}_{~~B]} .$$
Algebraic calculations show that   this is a    direct consequence of the motion equations. 

{\bf Local invariance}
$$0= \d \Pi _{IA} ~( \d  \lambda^{IA})
+  \Pi _{IJ} ~ (\d \lambda_K^{~~J} ~\omega^{IK}-\d\lambda^I_{~~K}~  \omega^{KJ}); $$

$$0=( \d \Pi _{AB}  -  \Pi _{JB}  ~\omega^J_{~~ A}
+  \Pi _{AJ}~  \omega_{B}^{~~J})~\d\lambda^{AB}.$$
This implies, taking antisymmetry into account,
$$  \d \Pi _{AB} =  \Pi _{JB}  ~\omega^J_{~~ A}
-  \Pi _{AJ}~  \omega_{B}^{~~J} =0,$$ also a consequence of the motion equations.
 
\section{Conclusions}
 
 We have expressed dynamics in terms of histories rather than dynamical variables. Although the set of histories has infinite dimensions,  we have defined a differential calculus on it which allowed us to perform variational calculus. A first benefit is that this formulation remains entirely covariant and does not require, for relativistic field theories (including \gr), the introduction of a time variable. In addition, it applies to the case where the histories (the \guill fields") are forms rather than functions, like for electromagnetism or the first order formulations of \gr. Such forms are treated as genuine objects and not decomposed into their components. 
 
 We have derived a very simple  expression for  the dynamical equations, analog to the usual \emph{Lagrange  equations}, but with an extended generality (covariance and treatment of forms).
We have  derived (generalizations of)  the Lagrangian, presymplectic and  symplectic forms, and shown the conservation of the latter on shell. We have also defined the conserved currents associated to symmetries. 

     All these quantities, as well as (the generalization of ) observables are well defined, although they are not usual functions or forms, but rather new mathematical entities  that we have called Hmaps, for which we have defined proper calculations. 
           Although an \emph{observable} is generally considered as a  scalar-valued function, it appears here (in particular in the case of \gr) as a   form-valued map. We consider   this generalization as a natural benefit of our approach since        any form provides a scalar  by integration over a submanifold  of adapted dimension. This provides a procedure  to extract  scalar observables form the   generalized (form-valued) observables, with the help of  intermediary  submanifolds. This corresponds for instance  to  what is done in Loop Quantum Gravity through the introduction of the Holonomy-Flux algebra. 

The  observables, as they are naturally  defined here, depend on   histories,   not on the configuration or phase space variables.
\footnote{In the Hamiltonian formalism \cite{HHH},  It is possible  reformulate them  as functions, \eg, on phase space}. Since, by construction,   they depend on the whole history, they   remain constant  during the evolution. For instance, they necessarily commute with the Hamiltonian and with the constraints, so that they correspond to the  \emph{complete  observables} in the sense of  \cite{RovelliPartialobservables,Dittrich} (see also\cite{WestmanSonego}).
 
In a following paper \cite{HHH}, one   defines a generalized Legendre transform and an explicitely covariant Hamiltonian historical dynamics.

 \begin{appendices}
\section{Historical Functionals and Multi-index notations}  \label{multiindex} \label{APPB}

We introduce a   multi-index notation for skew indices characterizing the components of forms : we write 
$\underline{\alpha}$ for the  antisymmetrized sequence $
\alpha_1 ~...~ \alpha_r $. In addition, we write   $\d ^{\underline{\alpha}}=\d x ^{\alpha_1}~...~\d x ^{\alpha_r}$ and
the multivector $e_{\underline{\alpha}}=(e_{\alpha_1}, ..., {\alpha_r})$ (wedge products assumed).

An      r-forms expands as 
$$c=c_{\underline{\mu}} ~\d ^{\underline{\mu}}   =c_{\mu_1...\mu_r}~\d x ^{\mu_1}...\d x ^{\mu_r}.$$
The notation extends to Hmaps  through 
   $$~F =F_{\underline{\alpha}} ~\d ^{\underline{\alpha}}   :  as  ~ 
   F_{\underline{\alpha}}(c,\gamma)=
(   F (c,\gamma))_{\underline{\alpha}}.$$
   
 Then we have    \eql{momentar}{
\pder{F}{ c}= \pder{F_{\underline{\alpha}}}{   c_{\underline{\mu} }}~   (\innerp{ e_{\underline{\mu}}}  \d ^ {\underline{\alpha}})  .
}  
In the case where a form $P$  is expanded in dual components: $P=P^{\underline{\mu}}~\Vol _{\underline{\mu}}$, with 
$\Vol _{\underline{\mu}}=\innerp{e _{\underline{\mu}}} \Vol$,  then   \eql{momentar}{
\pder{F}{ P}=    \epsilon^{\underline{\mu}\underline{\nu}}~
~   \pder   {F_{\underline{ \alpha}}}{ P^{\underline{\mu}}}~ (    \innerp{e_{\underline{\nu}}} \d ^{ \underline{ \alpha}}),     }
where
 $    \epsilon^{\underline{\mu}\underline{\nu}}\eqbydef     \epsilon^{ \mu_1...\mu_r~\nu_1... \nu_{n-r}}.$

The validity of these formulas implies some conditions on the    grades of the forms involved, that we always assume fullfilled. 

In the  Lagrangian case,
$\calL=\ell~\Vol$,\\ 
$$ \pder{\calL}{ c}= \pder{\ell}{  c_{\underline{\mu} }}~\Vol_{\underline{\mu} };~~~ \pder{\calL}{ (\d c)}= \pder{\ell}{    c_{\underline{\mu} ,\nu}}~\Vol_{\underline{\mu},\nu }.$$ 
\eql{DL}{\D \calL=\D c~\pder{\calL}{ c} +\D(\d c) ~\pder{\calL}{\d c } ~ ,}   
where antisymmetrization over all indices is assumed.

\end{appendices}

 \end{document}